\documentclass[a4paper,11pt]{article}
\usepackage{pos}

\title{Searching for charged Higgs bosons with flavor-changing couplings at the LHC}
\author*{Mohamed Krab}
\affiliation{Department of Physics, National Taiwan University,\\
Taipei 10617, Taiwan}

\emailAdd{mkrab@hep1.phys.ntu.edu.tw}

\abstract{We investigate the LHC discovery prospects for charged Higgs bosons in the general two Higgs doublet model (G2HDM) that has flavor-changing neutral Higgs (FCNH) couplings. 
The FCNH $\rho_{tc}$ coupling induces intriguing $H^+$ production processes $c\bar b \to H^+$, $cg \to bH^+$, and $\bar bg \to \bar cH^+$ without CKM suppression.
Sizable $\rho_{tc}$ can drive the disappearance of antimatter from the Universe. 
In this contribution, we promote the resonant $c\bar b \to H^+$ production, followed by the bosonic $H^+ \to W^+ H$ decay. Discovery could be a harbinger of G2HDM with FCNH~couplings, and perhaps shed light on the baryon asymmetry of the Universe.}

\FullConference{The Thirteenth Annual Large Hadron Collider Physics (LHCP2025)\\
5-9 May 2025\\
Taipei\\}

\begin{document}
\maketitle

\section{Introduction}
Remarkably, a second Higgs doublet provides a potential mechanism to explain the matter–antimatter asymmetry of the Universe. It also predicts the existence of additional Higgs bosons besides the 125~GeV state observed at the LHC~\cite{ATLAS:2012yve,CMS:2012qbp}---most notably, charged Higgs bosons. Discovering such states would confirm the nonminimal nature of the Higgs sector and perhaps shed light on the origin of the baryon asymmetry of the Universe.

We consider the general two Higgs doublet model (G2HDM), i.e., without imposing the usual $Z_2$~symmetry. 
Besides the fermion masses, the model allows for a second set of Yukawa couplings.
Flavor constraints on the flavor-changing neutral Higgs (FCNH) coupling $\rho_{tc}$ are not particularly strong~\cite{Crivellin:2013wna}.
Sizable $\rho_{tc}$ can drive~\cite{Fuyuto:2017ewj,Kanemura:2023juv} electroweak
baryogenesis to account for the disappearance of antimatter from the Universe.
Interestingly, $\rho_{tc}$ induces the production processes $c\bar b \to H^+$~\cite{Hou:2024ibt}, $cg \to bH^+$~\cite{Ghosh:2019exx,WIP:2025}, and $\bar b g \to \bar cH^+$~\cite{Hou:2024bzh} without the CKM suppression present in the 2HDM with softly broken $Z_2$ symmetry, such as the popular 2HDM type-II.

In this contribution, we suggest searching for the charged Higgs boson at the LHC via its resonant $c\bar b \to H^+$ production, followed by the bosonic $H^+ \to W^+ H$ decay. 
The scalar $H$ may decay dominantly into $t\bar c$ final state, leading eventually to same-sign dilepton signals. 
We perform a collider analysis at the LHC based on two representative benchmark points (BPs).
The analysis yields promising results for the LHC running at 14~TeV collision energy.

\section{Framework}
Assuming a {\it CP}-conserving Higgs sector, one can write the most general Higgs potential of the G2HDM in the Higgs basis as~~\cite{Davidson:2005cw,Hou:2017hiw}
\begin{align}
 V(\Phi,\Phi') &= \mu_{11}^2|\Phi|^2 + \mu_{22}^2|\Phi'|^2
    - (\mu_{12}^2\Phi^\dagger\Phi' + \rm{H.c.}) \nonumber\\
 & \,\quad + \frac{\eta_1}{2}|\Phi|^4 + \frac{\eta_2}{2}|\Phi'|^4
   + \eta_3|\Phi|^2|\Phi'|^2  + \eta_4 |\Phi^\dagger\Phi'|^2 \nonumber \\
 & \,\quad + \left[\frac{\eta_5}{2}(\Phi^\dagger\Phi')^2
   + \left(\eta_6 |\Phi|^2 + \eta_7|\Phi'|^2\right) \Phi^\dagger\Phi' + \rm{H.c.}\right],
%\label{pot}
\end{align}
where $\eta_i$s are the quartic couplings and are real. The vacuum expectation value $v$ comes from $\Phi$ through the minimization condition $\mu_{11}^2 = - \frac{1}{2}\eta_1 v^2$---while $\left\langle \Phi'\right\rangle = 0$, hence $\mu_{22}^2 > 0$. 
A second minimization condition, $\mu^2_{12} = \frac{1}{6}\eta_6 v^2$, eliminates $\mu_{12}^2$ as a parameter.

The G2HDM scalars $h$, $H$, $A$, and $H^+$ couple to fermions by~\cite{Davidson:2005cw}
\begin{align}
 %\mathcal{L}_Y = %\supset
 & - \frac{1}{\sqrt{2}} \sum_{f = u, d, \ell}
       \bar f_{i} \bigg[\big(\lambda^f_{ij} s_\gamma
       + \rho^f_{ij} c_\gamma\big) + \big(\lambda^f_{ij} c_\gamma - \rho^f_{ij} s_\gamma\big)H -i\,{\rm sgn}(Q_f) \rho^f_{ij} A\bigg]  R f_{j}\nonumber \\
 & \, \ \qquad\qquad - \bar{u}_i\big[(V\rho^d)_{ij} R - (\rho^{u\dagger}V)_{ij} L\big]d_j H^+ 
  - \bar{\nu}_i\rho^\ell_{ij} R \ell_j H^+ +{\rm H.c.},
\label{Lyuk}
\end{align}
where $i,j = 1\text{--}3$ are generation indices, 
$L,R = (1\mp\gamma_5)/2$, $c_\gamma \equiv \cos\gamma$ denotes the mixing angle between the two {\it CP}-even scalars $h$ and $H$ (the angle $\gamma$ corresponds to $\beta-\alpha$ in Ref.~\cite{Davidson:2005cw}), $s_\gamma \equiv \sin\gamma$, and $V$ is the CKM matrix. 
In the alignment limit of $c_\gamma \to 0$, $h$ couples diagonally to fermions, while $H$ couples through the extra Yukawa couplings $\rho^f_{ij}$. 
We note that $A$ and $H^+$ couplings are independent of $\gamma$. In what follows, we drop the superscript $f$. For simplicity, we turn off all $\rho_{ij}$, except the extra top Yukawa couplings $\rho_{tc}$ and $\rho_{tt}$. 

We focus on the $c\bar b \to H^+ \to W^+ H$ process, where $\rho_{tc}$ induces the $c\bar b \to H^+$ production without CKM suppression, as one can see from Eq.~\ref{Lyuk}, while the $H^+ \to W^+ H$ decay arises through
\begin{equation}
    i\frac{g}{2}s_\gamma(H\partial^\mu H^+ - H^+\partial^\mu H)W^-_\mu + \rm{H.c.},
\end{equation}
where $g$ is the $SU(2)_L$ gauge coupling. 
In the alignment limit where $h = h_{\rm{SM}}$ ($s_\gamma \simeq 1$), the coupling $H^+W^- H$ is maximized, while the coupling $H^+W^- h$ vanishes. 
The coupling $H^+W^- A$ is independent of the mixing angle $\gamma$, and is therefore unaffected by the alignment.
Here, we investigate the flavor-changing decay $H \to t\bar c$. 
For simplicity, we set $s_\gamma = 1$ and assume $m_A \simeq m_{H^+}$ (to satisfy electroweak precision constraints). Two BPs are selected for illustration (see Table~\ref{table:BPs}).

\begin{table}[t]
	\centering
	{\begin{tabular}{l c c c c  c  c c  c c} %| c| c
			\hline\hline
			% &&&&&&&&& \\ %&&
			 & %$\eta_1$ &
			$\eta_2$ & $\eta_3$ & $\eta_4$ & $\eta_5$ %& $\, \eta_6 \,$
			& $\eta_7$ & $m_{H}$ & $m_A$ & $m_{H^+}$ & ${\mu_{22}^2/v^2}$  \\
			\hline
			BP1 & %0.26 &
			1.40 & 2.00 & {$-0.82$} & {$-0.82$} & %0 &
			{$-0.55$} & 200 & 300 & 300 & 0.49 \\
			BP2 & %0.26 &
			2.88 & 4.75 & {$-2.64$} & {$-2.64$} & %0 &
			\,\,\,~0.16 & 300 & 500 & 500 & 1.75 \\
			%   &&&&&&&&& \\ %&&
			\hline\hline
			%\hline
	\end{tabular}}
	\caption{Our selected BPs (all masses in GeV). For BP1, $\rho_{tc} = \rho_{tt} = 0.1$, while for BP2, $\rho_{tc} = 0.3,\,\rho_{tt} = 0.5$.}
	\label{table:BPs}
\end{table} 
\begin{figure*}[b]
	\centering
	\includegraphics[scale=0.5]{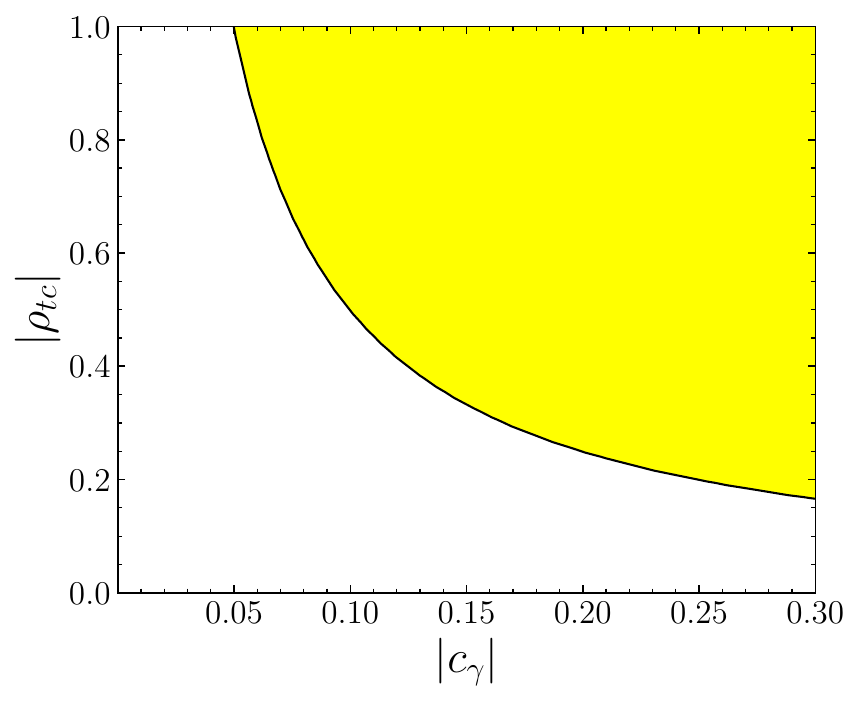}
	\caption{Exclusion bound from $t \to ch$ search in the $|c_\gamma|$-$|\rho_{tc}|$ plane.}
	\label{fig:tch-limit}
\end{figure*}  

\section{Constraints on $\rho_{tc}$}
ATLAS~\cite{ATLAS:2024mih} and CMS~\cite{CMS:2024ubt} searches for the flavor-changing $t \to ch$ decay set stringent limits on ${\cal B}(t \to c h) \leq 3.4 \times 10^{-4}$~\cite{ATLAS:2024mih} and ${\cal B}(t \to c h) \leq 3.7 \times 10^{-4}$~\cite{CMS:2024ubt}, respectively.
These limits place strong constraints on the FCNH coupling $\rho_{tc}$.
The exclusion bounds in the $|c_\gamma|$-$|\rho_{tc}|$ plane are depicted in Fig.~\ref{fig:tch-limit}.
LHC direct searches for heavy Higgs bosons with flavor-changing couplings place further constraints on $\rho_{tc}$. In the absence of interference between $H$ and $A$, CMS~\cite{CMS:2023xpx} set no exclusion limits on $m_H$ or $m_A$ for a coupling value of $\rho_{tc} = 0.4$. When interference is taken into account, still assuming $\rho_{tc} = 0.4$, the exclusion limits reach $m_H = 290$~GeV and $m_A = 340$~GeV. For a larger coupling, $\rho_{tc} = 1.0$, the limits extend further, excluding $m_H$ up to 760~GeV and $m_A$ up to 810~GeV at 95\% confidence level (CL)~\cite{CMS:2023xpx}.
ATLAS~\cite{ATLAS:2023tlp} excludes masses of a heavy Higgs boson $m_H$ between 200 and 620~GeV at 95\% CL for couplings $\rho_{tt} = 0.4$, $\rho_{tc} = 0.2$, and $\rho_{tu} = 0.2$. 
These limits become less stringent for very small or vanishing $\rho_{tu}$~\cite{ATLAS:2023tlp}.
Flavor constraints on $\rho_{tc}$ are weak~\cite{Crivellin:2013wna}. 
Constraints on $\rho_{tt}$ are discussed in Ref.~\cite{Hou:2025tjp}. 

\section{Collider Analysis}
Our same-sign dilepton signature is defined as two leptons with the same charge and at least two jets, at least one of which is tagged as a $b$-jet, plus missing transverse energy. The main SM backgrounds arise from $tW$, $t\bar tW$, $WZ$, $tZj$, $t\bar tZ$, $t\bar th$, $ZZ$, and $4t$ processes. 
Additionally, a charge-flip (or $Q$-flip) background may contribute when the charge of a lepton is misidentified. Fake backgrounds can also arise when jets are misidentified as leptons.
These backgrounds come mainly from $t\bar t + \rm{jets}$ process~\cite{CMS:2023ftu,ATLAS:2016dlg,Alvarez:2016nrz} and are estimated using mis-identification probabilities of $\epsilon_{\rm{Qflip}}=10^{-3}$~\cite{ATLAS:2016dlg,Alvarez:2016nrz,CMS:2019rvj} and $\epsilon_{\rm{fake}}=10^{-4}$~\cite{ATLAS:2016dlg,Alvarez:2016nrz}.
The ``non-prompt'' backgrounds are not included in this analysis, as they are not properly modeled in Monte Carlo simulations.

We use \texttt{MadGraph5\_aMC@NLO}~\cite{Alwall:2014hca} to generate the leading-order (LO) signal and background events at $\sqrt{s} = 14~\rm{TeV}$. The events are passed to \texttt{Pythia-8.2}~\cite{Sjostrand:2014zea} for showering and hadronization, and then to~\texttt{Delphes-3.5.0}~\cite{deFavereau:2013fsa} for detector simulation.
We include one additional jet for the signal, ${t\bar t}W$, ${t\bar t}Z$, $tW$, $WZ$, and $ZZ$ background events, and two additional jets for $t\bar t+\rm{jets}$, using the MLM matching scheme~\cite{Alwall:2007fs}. The $tZj$, ${t\bar t}h$, and $4t$ backgrounds are estimated without additional jets. 
The LO background cross sections are rescaled by the appropriate $K$-factor~\cite{Hou:2024ibt} to account for higher-order QCD corrections, while signal cross sections are kept at LO. 

For event selection, we require a minimum of two jets, of which at least one is $b$-tagged, with $p_T > 20$ GeV and $\left|\eta\right|<2.5$, exactly two same-sign leptons (electron or muon), with the leading (subleading) lepton satisfying $p_T > 25$ ($20$) GeV with $\left|\eta\right|<2.5$. 
A separation of $\Delta R > 0.4$ is required between the two selected leptons, as well as between any lepton and any jet. Additionally, events with missing energy $E^{\rm{miss}}_{T} > 35$~GeV, and a scalar $p_T$ sum of all jets and the two same-sign leptons ($H_T$) less than 400 GeV are selected.
It should be noted that the $H_T$ selection cut is chosen to maximize the signal significance.
\begin{figure*}[t]
	\centering
	\includegraphics[scale=0.5]{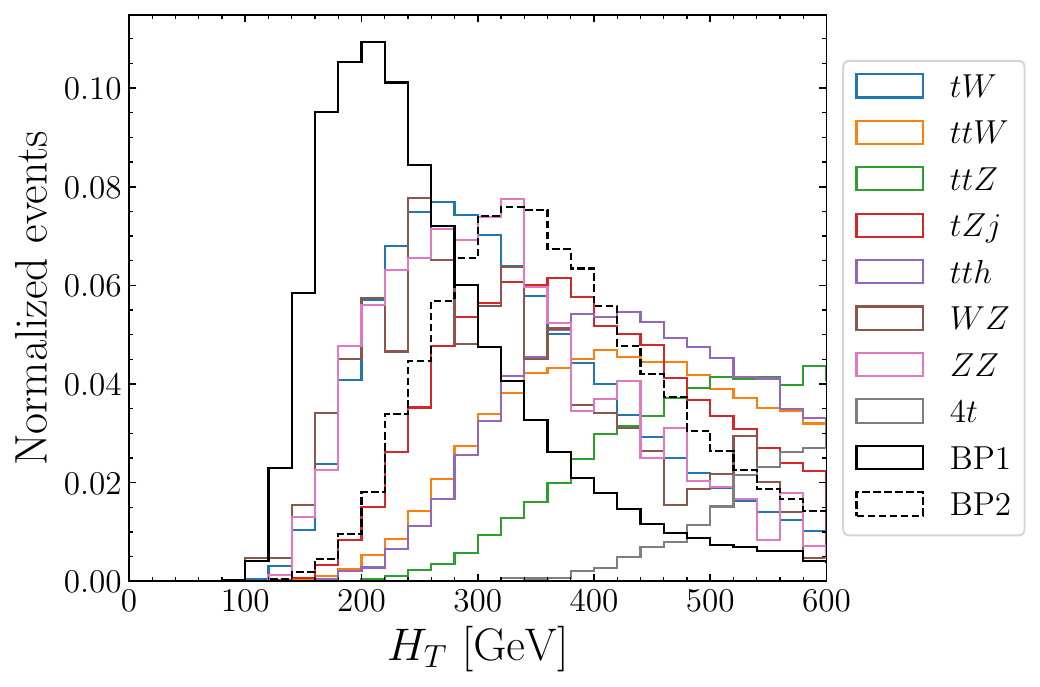}
	\caption{The normalized $H_T$ distribution for the signal and background processes.}
	\label{fig:HT}
\end{figure*}  
We plot in Fig.~\ref{fig:HT} the normalized $H_T$ distribution before the selection cuts. 
The signal and background cross sections after the selection cuts are given in Table~\ref{table:xsASC}. 
The $4t$, charge-flip and fake backgrounds contribute less than 1\% after the selection cuts. 

We estimate our signal sensitivity using~\cite{Kumar:2015tna}
\begin{equation}
	z = \sqrt{2\left[(S+B)\ln\left(\frac{(S+B)\left(B+\Delta_B^2\right)}{B^2 + (S+B)\Delta_B^2}\right)-\frac{B^2}{\Delta_B^2} + \ln\left(1+ \frac{\Delta_B^2 S}{B(B+\Delta_B^2)}\right) \right]},
\end{equation}
with $\Delta_B = \epsilon B$, where $S$ ($B$) is the number of signal (background) events, and $\epsilon$ refers to the systematic uncertainty in the background estimation.
Assuming $\epsilon = 10\%$, BP1 (BP2) yields a significance of $\sim 7.4\sigma$ ($8.8\sigma$) at 300~fb$^{-1}$.
The significances as a function of luminosity for the systematic uncertainties of 5\%, 10\% and 20\% can be found in Fig.~3 of Ref.~\cite{Hou:2024ibt}.

\begin{table}[t]
	\centering
	{ \begin{tabular}{l c c c c c c c c} %| c| c 
			\hline\hline
			% &&&&&&&&& \\ %&&
	Process & BP1~(BP2) & $tW$ & $t\bar tW$ & $WZ$ & $tZj$ & $t\bar tZ$ & $tth$ & $ZZ$ \\	\hline
	Cross section & 3.72~(4.62) & 1.61 & 1.09 & 0.54 & 0.40 & 0.10 & 0.05 & 0.02 \\
			\hline\hline
			%\hline
	\end{tabular}}
	\caption{Signal and background cross sections (fb) after selection cuts.}
	\label{table:xsASC}
\end{table} 

\section{Conclusion}
We suggest searching for charged Higgs bosons with FCNH couplings. Induced by the FCNH $\rho_{tc}$ coupling, $H^+$ can be produced resonantly at tree level via $c\bar b \to H^+$ and decayed to $W^+H$. The scalar $H$ may decay dominantly into $t\bar c$ final state, giving rise to dilepton signals. 
We performed a detector-level Monte Carlo analysis at the 14~TeV LHC and found that discovery could be achieved at this energy stage. 
One would measure the FCNH coupling $\rho_{tc}$ and may test our understanding of the baryon asymmetry of the Universe.

\section*{Acknowledgment}
This work is supported by the National Science and Technology Council of Taiwan under grant No.~113-2639-M-002-006-ASP.
We thank Wei-Shu Hou for the pleasant collaboration.

\bibliographystyle{JHEP}
\bibliography{mybib}

\providecommand{\href}[2]{#2}\begingroup\raggedright\begin{thebibliography}{10}

\bibitem{ATLAS:2012yve}
{\bf ATLAS} Collaboration, G.~Aad et~al., {\it {Observation of a new particle
  in the search for the Standard Model Higgs boson with the ATLAS detector at
  the LHC}},  {\em Phys. Lett. B} {\bf 716} (2012) 1--29,
  [\href{http://arxiv.org/abs/1207.7214}{{\tt arXiv:1207.7214}}].

\bibitem{CMS:2012qbp}
{\bf CMS} Collaboration, S.~Chatrchyan et~al., {\it {Observation of a New Boson
  at a Mass of 125 GeV with the CMS Experiment at the LHC}},  {\em Phys. Lett.
  B} {\bf 716} (2012) 30--61, [\href{http://arxiv.org/abs/1207.7235}{{\tt
  arXiv:1207.7235}}].

\bibitem{Crivellin:2013wna}
A.~Crivellin, A.~Kokulu, and C.~Greub, {\it {Flavor-phenomenology of
  two-Higgs-doublet models with generic Yukawa structure}},  {\em Phys. Rev. D}
  {\bf 87} (2013), no.~9 094031, [\href{http://arxiv.org/abs/1303.5877}{{\tt
  arXiv:1303.5877}}].

\bibitem{Fuyuto:2017ewj}
K.~Fuyuto, W.-S. Hou, and E.~Senaha, {\it {Electroweak baryogenesis driven by
  extra top Yukawa couplings}},  {\em Phys. Lett. B} {\bf 776} (2018) 402--406,
  [\href{http://arxiv.org/abs/1705.05034}{{\tt arXiv:1705.05034}}].

\bibitem{Kanemura:2023juv}
S.~Kanemura and Y.~Mura, {\it {Electroweak baryogenesis via top-charm mixing}},
   {\em JHEP} {\bf 09} (2023) 153, [\href{http://arxiv.org/abs/2303.11252}{{\tt
  arXiv:2303.11252}}].

\bibitem{Hou:2024ibt}
W.-S. Hou and M.~Krab, {\it {Searching for resonant flavor-changing charged
  Higgs production at the LHC}},  {\em Phys. Rev. D} {\bf 111} (2025), no.~3
  L031701, [\href{http://arxiv.org/abs/2409.18474}{{\tt arXiv:2409.18474}}].

\bibitem{Ghosh:2019exx}
D.~K. Ghosh, W.-S. Hou, and T.~Modak, {\it {Sub-TeV $H^+$ Boson Production as
  Probe of Extra Top Yukawa Couplings}},  {\em Phys. Rev. Lett.} {\bf 125}
  (2020), no.~22 221801, [\href{http://arxiv.org/abs/1912.10613}{{\tt
  arXiv:1912.10613}}].

\bibitem{WIP:2025}
C.~Fang, W.-S. Hou, C.~Kao, and M.~Krab, {\it {Enhanced Charged Higgs Signal at
  the LHC}},  {in preparation}.

\bibitem{Hou:2024bzh}
W.-S. Hou and M.~Krab, {\it {Reconstructing the general 2HDM charged Higgs
  boson at the LHC}},  {\em Phys. Rev. D} {\bf 110} (2024), no.~1 L011702,
  [\href{http://arxiv.org/abs/2405.19190}{{\tt arXiv:2405.19190}}].

\bibitem{Davidson:2005cw}
S.~Davidson and H.~E. Haber, {\it {Basis-independent methods for the
  two-Higgs-doublet model}},  {\em Phys. Rev. D} {\bf 72} (2005) 035004,
  [\href{http://arxiv.org/abs/hep-ph/0504050}{{\tt hep-ph/0504050}}]. [Erratum:
  Phys.Rev.D 72, 099902 (2005)].

\bibitem{Hou:2017hiw}
W.-S. Hou and M.~Kikuchi, {\it {Approximate Alignment in Two Higgs Doublet
  Model with Extra Yukawa Couplings}},  {\em EPL} {\bf 123} (2018), no.~1
  11001, [\href{http://arxiv.org/abs/1706.07694}{{\tt arXiv:1706.07694}}].

\bibitem{ATLAS:2024mih}
{\bf ATLAS} Collaboration, G.~Aad et~al., {\it {Search for flavour-changing
  neutral-current couplings between the top quark and the Higgs boson in
  multi-lepton final states in 13~TeV pp collisions with the ATLAS detector}},
  {\em Eur. Phys. J. C} {\bf 84} (2024), no.~7 757,
  [\href{http://arxiv.org/abs/2404.02123}{{\tt arXiv:2404.02123}}].

\bibitem{CMS:2024ubt}
{\bf CMS} Collaboration, A.~Hayrapetyan et~al., {\it {Search for
  flavor-changing neutral current interactions of the top quark mediated by a
  Higgs boson in proton-proton collisions at 13 TeV}},  {\em Phys. Rev. D} {\bf
  112} (2025), no.~3 032008, [\href{http://arxiv.org/abs/2407.15172}{{\tt
  arXiv:2407.15172}}].

\bibitem{CMS:2023xpx}
{\bf CMS} Collaboration, A.~Hayrapetyan et~al., {\it {Search for new Higgs
  bosons via same-sign top quark pair production in association with a jet in
  proton-proton collisions at $\sqrt{s}$ = 13 TeV}},  {\em Phys. Lett. B} {\bf
  850} (2024) 138478, [\href{http://arxiv.org/abs/2311.03261}{{\tt
  arXiv:2311.03261}}].

\bibitem{ATLAS:2023tlp}
{\bf ATLAS} Collaboration, G.~Aad et~al., {\it {Search for heavy Higgs bosons
  with flavour-violating couplings in multi-lepton plus b-jets final states in
  pp collisions at 13 TeV with the ATLAS detector}},  {\em JHEP} {\bf 12}
  (2023) 081, [\href{http://arxiv.org/abs/2307.14759}{{\tt arXiv:2307.14759}}].

\bibitem{Hou:2025tjp}
W.-S. Hou and M.~Krab, {\it {Probing the general 2HDM with flavor violation
  through A{\textrightarrow}ZH}},  {\em Phys. Rev. D} {\bf 111} (2025), no.~11
  115036, [\href{http://arxiv.org/abs/2503.23133}{{\tt arXiv:2503.23133}}].

\bibitem{CMS:2023ftu}
{\bf CMS} Collaboration, A.~Hayrapetyan et~al., {\it {Observation of four top
  quark production in proton-proton collisions at s=13TeV}},  {\em Phys. Lett.
  B} {\bf 847} (2023) 138290, [\href{http://arxiv.org/abs/2305.13439}{{\tt
  arXiv:2305.13439}}].

\bibitem{ATLAS:2016dlg}
{\bf ATLAS} Collaboration, G.~Aad et~al., {\it {Search for supersymmetry at
  $\sqrt{s}=13$ TeV in final states with jets and two same-sign leptons or
  three leptons with the ATLAS detector}},  {\em Eur. Phys. J. C} {\bf 76}
  (2016), no.~5 259, [\href{http://arxiv.org/abs/1602.09058}{{\tt
  arXiv:1602.09058}}].

\bibitem{Alvarez:2016nrz}
E.~Alvarez, D.~A. Faroughy, J.~F. Kamenik, R.~Morales, and A.~Szynkman, {\it
  {Four tops for LHC}},  {\em Nucl. Phys. B} {\bf 915} (2017) 19--43,
  [\href{http://arxiv.org/abs/1611.05032}{{\tt arXiv:1611.05032}}].

\bibitem{CMS:2019rvj}
{\bf CMS} Collaboration, A.~M. Sirunyan et~al., {\it {Search for production of
  four top quarks in final states with same-sign or multiple leptons in
  proton-proton collisions at $\sqrt{s}=$ 13 TeV}},  {\em Eur. Phys. J. C} {\bf
  80} (2020), no.~2 75, [\href{http://arxiv.org/abs/1908.06463}{{\tt
  arXiv:1908.06463}}].

\bibitem{Alwall:2014hca}
J.~Alwall, R.~Frederix, S.~Frixione, V.~Hirschi, F.~Maltoni, O.~Mattelaer,
  H.~S. Shao, T.~Stelzer, P.~Torrielli, and M.~Zaro, {\it {The automated
  computation of tree-level and next-to-leading order differential cross
  sections, and their matching to parton shower simulations}},  {\em JHEP} {\bf
  07} (2014) 079, [\href{http://arxiv.org/abs/1405.0301}{{\tt
  arXiv:1405.0301}}].

\bibitem{Sjostrand:2014zea}
T.~Sj{\"o}strand, S.~Ask, J.~R. Christiansen, R.~Corke, N.~Desai, P.~Ilten,
  S.~Mrenna, S.~Prestel, C.~O. Rasmussen, and P.~Z. Skands, {\it {An
  introduction to PYTHIA 8.2}},  {\em Comput. Phys. Commun.} {\bf 191} (2015)
  159--177, [\href{http://arxiv.org/abs/1410.3012}{{\tt arXiv:1410.3012}}].

\bibitem{deFavereau:2013fsa}
{\bf DELPHES 3} Collaboration, J.~de~Favereau, C.~Delaere, P.~Demin,
  A.~Giammanco, V.~Lema{\^\i}tre, A.~Mertens, and M.~Selvaggi, {\it {DELPHES 3,
  A modular framework for fast simulation of a generic collider experiment}},
  {\em JHEP} {\bf 02} (2014) 057, [\href{http://arxiv.org/abs/1307.6346}{{\tt
  arXiv:1307.6346}}].

\bibitem{Alwall:2007fs}
J.~Alwall et~al., {\it {Comparative study of various algorithms for the merging
  of parton showers and matrix elements in hadronic collisions}},  {\em Eur.
  Phys. J. C} {\bf 53} (2008) 473--500,
  [\href{http://arxiv.org/abs/0706.2569}{{\tt arXiv:0706.2569}}].

\bibitem{Kumar:2015tna}
N.~Kumar and S.~P. Martin, {\it {Vectorlike Leptons at the Large Hadron
  Collider}},  {\em Phys. Rev. D} {\bf 92} (2015), no.~11 115018,
  [\href{http://arxiv.org/abs/1510.03456}{{\tt arXiv:1510.03456}}].

\end{thebibliography}\endgroup

\end{document}